\documentclass{article}

\usepackage{amsmath}

\usepackage{amsmath}
\usepackage{algorithmic}
\usepackage[table]{xcolor}
\usepackage{graphicx} 
\usepackage{todonotes}
\usepackage{hyperref}

\usepackage[utf8]{inputenc} 
\usepackage[T1]{fontenc}    
\usepackage{hyperref}       
\usepackage{url}            
\usepackage{booktabs}       
\usepackage{amsfonts}       
\usepackage{nicefrac}       
\usepackage{microtype}      
\usepackage{xcolor}         

\usepackage{microtype}
\usepackage{graphicx}
\usepackage{subfigure}
\usepackage{booktabs} 

\usepackage{eso-pic}
\usepackage{xparse}

\newif\ifartifact

\usepackage[accepted]{mlsys2025}

\mlsystitlerunning{AdaParse: Adaptive Parallel PDF Parsing}

\def\showcomments{} 

\ifdefined\showcomments
    \newcommand{\note}[1]{{\textcolor{red}{\emph{#1}}}}
    \newcommand{\robert}[1]{\textcolor{red}{Robert: #1}}
    \newcommand{\carlo}[1]{\textcolor{blue}{Carlo: #1}}
    \newcommand{\ian}[1]{\textcolor{green}{Ian: #1}}

\else
    \newcommand{\note}[1]{}
    \newcommand{\robert}[1]{}
    \newcommand{\carlo}[1]{}
    \newcommand{\ian}[1]{}
\fi

\ifartifact
\newlength{\badgewidth}
\newlength{\badgegap}
\newcommand{\badgeList}{}

\NewDocumentCommand{\addTopRightBadge}{O{} m}{%
\gappto{\badgeList}{\href{#1}{\includegraphics[width=\badgewidth]{#2}}\hspace{\badgegap}}%
}

\newcommand{\placeTopRightBadges}{%
\AddToShipoutPictureBG*{%
\put(\LenToUnit{\paperwidth - 1.5cm - \badgewidth},\LenToUnit{\paperheight - 2cm}){%
\makebox[0pt][r]{\badgeList}%
}%
}%
}

\addTopRightBadge{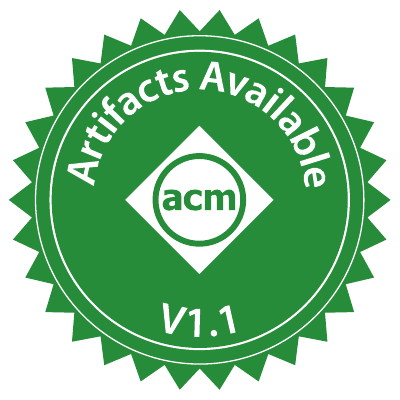}
\addTopRightBadge{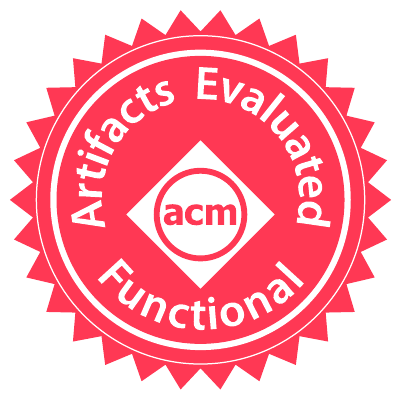}

\placeTopRightBadges
\fi

\begin{document}

\twocolumn[
\mlsystitle{AdaParse: An Adaptive Parallel PDF Parsing\\ and Resource Scaling Engine}
\title{}

\mlsyssetsymbol{equal}{*}

\begin{mlsysauthorlist}
\mlsysauthor{Carlo~Siebenschuh}{uc,anl} 
\mlsysauthor{Kyle~Hippe}{uc,anl} 
\mlsysauthor{Ozan~Gokdemir}{uc,anl} 
\mlsysauthor{Alexander~Brace}{uc,anl} 
\mlsysauthor{Arham~Khan}{uc}  
\mlsysauthor{Khalid~Hossain}{anl}  
\mlsysauthor{Yadu~Babuji}{anl}  
\mlsysauthor{Nicholas~Chia}{anl}  
\mlsysauthor{Venkatram~Vishwanath}{anl}  
\mlsysauthor{Rick~Stevens}{uc,anl}  
\mlsysauthor{Arvind~Ramanathan}{uc,anl} 
\mlsysauthor{Ian Foster}{uc,anl} 
\mlsysauthor{Robert~Underwood}{anl} 
\end{mlsysauthorlist}

\mlsysaffiliation{uc}{Department of Computer Science, University of Chicago, Chicago, Illinois, USA}
\mlsysaffiliation{anl}{Argonne National Laboratory, Lemont, Illinois, USA}

\mlsyscorrespondingauthor{Carlo~Siebenschuh}{siebenschuh@uchicago.edu}

\mlsyskeywords{Parsing, DPO}

\vskip 0.3in


\begin{abstract}
    Language models for scientific tasks are trained on text from scientific publications---most distributed as PDFs that require parsing. 
    PDF parsing approaches range from inexpensive heuristics (for simple documents) to computationally intensive ML‑driven systems (for complex or degraded ones). The choice of the ``best'' parser for a particular document depends on 1) its computational cost and 2) the accuracy of its output.
    To address these issues, we introduce an Adaptive Parallel PDF Parsing and Resource Scaling Engine (AdaParse), a data-driven strategy for assigning an appropriate parser to each document. We enlist scientists to select preferred parser outputs and incorporate this information through direct preference optimization (DPO) into AdaParse, thereby aligning its selection process with human judgment. AdaParse then incorporates hardware requirements and (aligned) predicted accuracy of each parser to orchestrate computational resources efficiently for large-scale parsing campaigns. We demonstrate that AdaParse, when compared to state-of-the-art parsers, improves throughput by 17$\times$ while still achieving comparable accuracy (actually, 0.2\% better) on a benchmark set of 1000 scientific documents. AdaParse's combination of high accuracy and parallel scalability makes it feasible to parse large-scale scientific document corpora to support the development of high-quality, trillion-token-scale text datasets. 
    \newline The implementation is available at \url{https://github.com/7shoe/AdaParse/}.
\end{abstract}
]

\printAffiliationsAndNotice{}  

\section{Introduction}

The great wealth of information stored in the scientific literature and 
the successes of large language models (LLMs) motivate efforts to train science-specialized LLMs on scientific documents \cite{beltagy2019scibert,taylor2022galactica}.
However such efforts require immense amounts of text for training \cite{chowdhery_palm_2022,li2024scilitllm}, and much of it is represented in Portable Document Format (PDF).
Exploiting this text requires correctly parsing information from PDFs, which is challenging due to their print-focused layout-based structure that is not designed for machine readability \cite{coulon2023machine}.
For complex PDFs, lightweight parsers often extract text swiftly but incorrectly, introducing artifacts that degrade the performance of LLMs trained on it. Mitigating adverse effects requires substantially more training data to achieve the same final LLM performance, further exacerbating the challenge \cite{sorscher_beyond_2023}.

As we will show, state-of-the-art high-quality parsing software is extremely computationally demanding, being able to parse only 1--2 PDF/s on a node with 4~A100 GPUs, making them impractical for datasets of hundreds of millions of scientific papers (\autoref{fig:motivateadaptive}).
Unsurprisingly, existing datasets of scientific tokens suitable for LLM training are modest in size (e.g., the popular Dolma dataset contains only 70B tokens from scientific sources \cite{soldaini_dolma_2024_correct_caps}) and often suffer from poor parse quality \cite{bast2017benchmark}.
Thus \textit{accurate} and \textit{efficient} PDF parsing is a central problem for those seeking to build high-quality AI-based scientific assistants and similar tools. 

However, not all is lost. 
As we will show, many ``simpler'' documents can be parsed with lightweight tools orders of magnitude faster with similar (or even improved) output quality as compared to their compute-intensive counterparts.
We leverage this fact to develop an adaptive parsing strategy that invokes lightweight parsers on simpler documents while reserving high-quality parsers for those deemed complex---thus deploying the most promising parser for each particular PDF in a way that balances competing demands for accuracy and throughput.
We show that this approach can greatly improve overall goodput as measured by \textit{accepted} textual tokens generated per resource unit.

To realize these benefits, this work makes the following contributions:
\begin{itemize}
\item A comprehensive benchmark to assess parser performance characteristics, conducted on 25,000 PDFs from across disciplines and publishers, including an assessment of how human perception and parser output quality align with commonly used metrics.
\item A predictive algorithm that, for any PDF, selects the parser most likely to yield accurate output, adapting to document attributes and user preferences.
\item The integrated design of batching, prefetching, parallel execution, and scheduling to realize adaptive parsing of PDFs for high throughput and quality on leadership-class HPC systems.
\end{itemize}

The remainder of the paper is organized as follows:
In Section~\ref{sec:background}, we describe parsing challenges and failure modes, how quality is compared between parsers, and why PDF parsing presents a non-trivial parallel and distributed systems problem.
Next, in Section~\ref{sec:related}, we describe the various classes of parsers and how they can be used in parallel workflows to parse PDFs \textit{en masse}.
After that, we formulate our task of producing high-quality text output as an optimization problem in Section~\ref{sec:problem}.
We then provide an overview of our system in Section~\ref{sec:design} and present the key optimizations that we employ to achieve both high throughput and high-quality text.
We discuss our experimental methodology in Section~\ref{sec:methods}.
Finally, we present our evaluation in Section~\ref{sec:evaluation}, followed by conclusions and future work in Section~\ref{sec:conclusions}.

\section{Background}\label{sec:background}

\subsection{Challenges in PDF Parsing}
\begin{figure}[h]
    \centering
    \includegraphics[width=0.95\columnwidth, trim=150 90 100 15, clip]{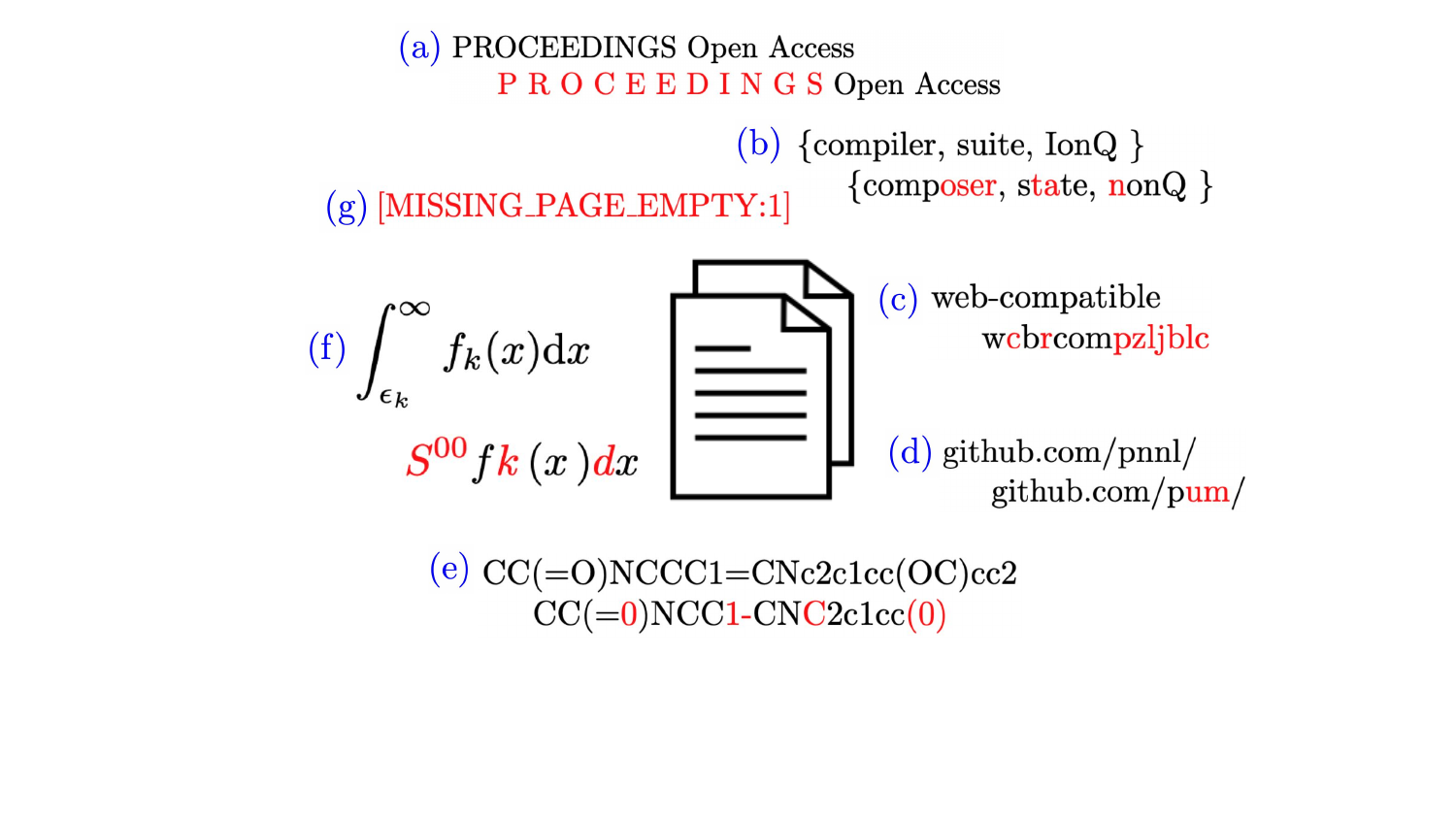}
    \caption{Failure modes of PDF parsers, (a) whitespace injection, (b) word substitution, (c) character scrambling, (d) character substitution, (e) corrupted SMILES, (f) LaTeX to plaintext conversion, (g) document page dropped.}
    \label{fig:failuremodes}
\end{figure}

Document parsers can fail to produce accurate text output in a variety of ways.
As illustrated in Figure~\ref{fig:failuremodes}, failures can include introducing whitespace, substituting words, scrambling characters or words, corrupting identifiers or references, or even dropping entire pages. Notably, such errors are not confined to lightweight parsers; even sophisticated parsing software can encounter them. In fact, we have found that the most severe failure mode---dropping an entire page---occurs with the parser that otherwise delivers the most accurate results.

These failures are driven by the fact that the PDF is layout-driven: it is designed to provide versatility in achieving a desired visual appearance. This versatility means that parsing even a single PDF document can be challenging. Born-digital PDFs can contain diverse elements such as figures, tables, and rich media like videos \cite{correa2017unleashing}. They may even hold hidden information or malware \cite{kuribayashi2021stealthpdf,singh2020malware}. Scanned documents, on the other hand, suffer from significant quality degradation, complicating content extraction \cite{mujumdar2019simultaneous}. This diversity virtually rules out the development of a universal parsing strategy, as a one-size-fits-all parser would either be too simplistic to handle complex cases or inefficient in parsing simpler ones. This situation necessitates the use of an \textit{adaptive} parsing strategy able to address each PDF individually. Yet naïvely applying each available parser to a given document and selecting the output that appears the most accurate is infeasible. Thus, parser selection must be based on easily available data and form a prediction on it \cite{ravi2008automatic}.  

\subsection{Evaluation of Parsing Accuracy}
A key obstacle to evaluating PDF parsers is the lack of a quality metric for measuring their output against groundtruth text. In the absence of a universal accuracy measure that comprehensively captures the similarity between long-form texts---accounting for syntax, spelling errors, and scientific content---several proxy metrics are used.

Traditional metrics assess the similarity between parser output and groundtruth text on a character level. For example, the Levenshtein distance reports the minimum number of character edits required to transform one text into the other \cite{levenshtein1966binary}. Although straightforward to compute, it may poorly align with human perception of quality \cite{nerbonne1999edit}. Moreover, these routines can prove computationally prohibitive for ultra-long text sequences as encountered in parsed PDF text. Moreover, scientific (in-)accuracy goes beyond character errors that may prove subtle but deadly. For example, while the edit distance between “hyperthyroidism” and “hypothyroidism” is just two, implying a normalized similarity of 86.7\%, the treatments for these conditions are opposites. Similarly, changes in character capitalization can turn the measure of acidity (pH) into the phenyl group (Ph).

Modern metrics such as BLEU (Bilingual Evaluation Understudy) \cite{bleu} and ROUGE (Recall-Oriented Understudy for Gisting Evaluation) \cite{rouge} set out to measure string similarity in a manner more aligned with human perception. These metrics are based on the number of matching n-grams and capture meaning across multiple words. Regardless, they may still fail to capture scientific meaning even when evaluating a single sentence. For instance, consider the following groundtruth text:
\begin{quotation}
\noindent
``\textit{The gravitational force between two masses is directly proportional to the product of their masses and inversely proportional to the square of the distance between them}.''
\end{quotation}
and the candidate text:
\begin{quotation}
\noindent
``\textit{The gravitational force inversely masses the proportional distance between two products and is directly proportional to the square of objects}.''
\end{quotation}
When evaluating under BLEU and ROUGE, we observe 0.32 and 0.82 respectively---indicating reasonable and high accuracy to the groundtruth text, despite the incoherent and factually erroneous candidate text. 

Furthermore, these metrics are designed to assess the sentence-length quality of neural translations rather than the multi-page parser output of scientific documents \cite{graham2015re}. The accumulation of seemingly small mistakes can result in text that appears to be of high quality but significantly distorts the intended insights. Finally, current parsers require hyperparameters that are only tacitly assumed fixed and can hardly be considered canonical \cite{post2018call}. While indicative of perceived text quality, these metrics are insufficient to thoroughly compare PDF parser output against the groundtruth text. 

\subsection{Datasets and Benchmarks}
Several datasets have been created to assess or improve PDF parsing technology \cite{jimeno2021icdar}. Early datasets focused on specific applications such as license plate or business card recognition \cite{bulan2017segmentation, saiga1993ocr}. Recent datasets, on the other hand, cater to specific document types, including handwritten notes, scanned documents, and layout-rich publications \cite{shaffi2021uthcd, jaume2019funsd, zhu2022docbed, paudel2024optimizing}. A recent survey indicates that scientific documents are exceptionally challenging for parsers \cite{adhikari2024comparative}.

S2ORC \cite{lo2019s2orc} is particularly relevant to scientific information parsing as it contains over 8 million full-text academic papers from diverse publishers. It is not suitable for this work, however, as groundtruth text was synthesized by a PDF parser (GROBID), and the resulting text/PDF pairs served as the training data of another (Nougat) \cite{blecher2023nougat_new}. Maintaining the integrity of our benchmark requires the use of text and PDF pairs that have not been incorporated into the training of neural networks deployed by PDF parsers. However, recent research continues to prioritize dataset size over the quality of annotations or access to closed-source content. As AI-driven parsing systems evolve, it becomes increasingly important that their evaluation undergoes equally rigorous, AI-level scrutiny to ensure reliability and accuracy \cite{paudel2024optimizing}.

\subsection{Parallel Systems}

For a given parser, runtime depends on the content of a PDF, which can include vector graphics, raster images, or even multimedia elements. Runtimes vary even more widely across parsing strategies, with some parsers employing large machine learning (ML) models to process documents line by line. Since it is a priori unknown what parser is best to handle a given PDF, the overall time to parse text is subject to a great deal of uncertainty.

Writing software to parse a single PDF is tedious, but scaling that software to 100 million PDFs is a considerable challenge for parallel and distributed systems. Parsing involves I/O-intensive workloads, where large batches of PDFs are read into memory and substantial text data is written to distributed storage. Heterogeneous PDFs lead to varying batch sizes and uneven processing times, complicating load balancing across distributed nodes. A resilient infrastructure is necessary to handle corrupted PDFs that may be present in datasets, and potential security risks can arise if PDFs contain malicious software. Additionally, indexing the parsed text is challenging, as the lack of consistency across documents complicates maintaining accurate metadata.

\section{Related Work}\label{sec:related}


\subsection{Parsers}
We can distinguish two classes of PDF parsing methods: text extraction and text recognition. Text recognition, in turn, includes both traditional optical character recognition (OCR) techniques and newer approaches that leverage modern ML models, such as Vision Transformers (ViT).

\subsubsection{Extraction}
Text extraction tools retrieve content directly from the textual layer embedded within a PDF. MuPDF, for example, is a high-performance extraction and rendering tool \cite{MuPDF}. Its Python binding PyMuPDF supports various input and output file types, such as LlamaIndex, a format tailored to LLM data curation. Another popular extraction tool, pypdf \cite{pypdf}, is a pure Python library.

Extraction tools are generally fast and language-agnostic, indiscriminately retrieving the entirety of text embedded within a document. However, they falter when text is either not embedded explicitly or is of poor quality. Text scrambling is sometimes employed by authors to obstruct extraction. Even if the text is embedded with good intentions, it may still be of low quality if initially inferred and attached by subpar text recognition software.

\subsubsection{Optical Character Recognition (OCR)}
Text recognition addresses these challenges by converting images of text into machine-readable formats. Optical character recognition employs computer vision techniques to transcribe characters line-by-line. OCR is commonly applied to scanned documents to create an explicit text layer. Numerous libraries support transformation of documents into searchable, structured text \cite{neudecker2021survey}.

Tesseract is an open-source OCR engine that has been refined over four decades, predating the PDF itself \cite{smith2007overview_new}. Now in its fifth version, it employs long short-term memory networks (LSTMs) to infer text sequences from image input. However, its ability to adapt these models to specific document corpora has been limited as training functionality is no longer supported. GROBID (GeneRation Of BIbliographic Data) is another tool that combines machine learning with text extraction to generate highly structured outputs \cite{GROBID}. GROBID natively provides some parallel parsing capabilities via multi-threading. It is particularly well-suited for scientific document parsing, offering features such as references, affiliations, and metadata extraction. GROBID is flexible, utilizing entity-specific ML models for bibliographic data extraction and large language models (LLMs) for text completion. It represents a growing trend toward blending classical OCR with modern ML tools \cite{lopez2009grobid}.

OCR is generally robust in parsing text from documents as it does not rely on an embedded text layer. However, OCR is computationally intensive, often operating orders of magnitude slower than text extraction tools. Throughput can be further reduced by the need for post-processing to properly format the extracted text \cite{nguyen2021survey}. Finally, OCR models often require training or calibration for optimal performance. Unsurprisingly, modern OCR implementations frequently rely on GPUs for improved efficiency \cite{du2020pp}.

\subsubsection{Vision Transformers (ViTs)}
Recent innovations have led to the development of Transformer-based neural architectures for OCR \cite{li2023trocr}. Such Vision Transformers (ViTs) learn to decode text from page images in an end-to-end manner. The Document Understanding Transformer (Donut) pioneered this approach for document text recognition, initially focusing on receipts \cite{kim2022ocr_new}. Nougat and $\mu$gat extended these capabilities to the parsing of scientific PDFs \cite{quattrini2024mu_new}. Marker further refines this approach through explicit layout detection that precedes parsing of individual document elements through texify \cite{marker_datalab}. 

Vision Transformers have shown significant promise in parsing scientific documents. They excel at navigating layout-dense PDF pages and are specifically trained to decode LaTeX equations. However, ViTs are highly compute-intensive at inference time, with their runtime scaling quadratically in the number of image patches. Even with vast datasets of PDF and groundtruth text pairs, along with the computational power required for training, their ability to generalize to unseen document types remains uncertain---particularly in the absence of properly held-out benchmark data.

\subsection{Adaptive Parsing}
A substantial body of research focuses on training and applying neural networks to tasks involving the prediction of various accuracy metrics. The use of data-driven models to predict a document parser's performance, or to select an appropriate parser through classification, is not an entirely novel concept. For instance, methods have been developed to estimate parser accuracy based on document metadata \cite{ravi2008automatic}. Other approaches have explored selective content parsing \cite{zuidema2007parsimonious} or training parser ensembles via bootstrapping \cite{steedman2003bootstrapping}. However, much of this earlier work centered on short text inputs (e.g., sentences), predating the rise of large language models and advances in the classification and regression of long-form text, which can now be leveraged to more effectively predict optimal parser candidates.

\subsection{Scientific Corpora}

Despite the wide range of datasets for training LLMs, there are just two major sources of scientific articles---PILE \cite{gao_pile_2020_correct_caps} (which contains, for example, ArXiV in \LaTeX \, form) and S2ORC \cite{lo2019s2orc}---that are used in the training of open LLMs.
However, these two sources can be vastly under-inclusive of scientific documents.  From obtaining access to collections such as the ACM Digital Library and comparing them to these sources, we have measured that as many as 80\% of scientific documents from publishers like ACM are not contained in PILE and S2ORC, presenting a gap that would be addressed through adaptive and high-quality PDF parsing.

The PILE dataset includes a subset called PhilPapers, which consists of academic PDFs parsed using Apache PDFBox \cite{gao_pile_2020_correct_caps}. Regardless, the amount of scientific content is relatively small and almost exclusively sourced from \LaTeX\ sources of ArXiV rather than PDFs. While parsing from LaTeX can be more reliable than parsing from PDFs, LaTeX sources are not accessible for most papers.

Semantic Scholar's S2ORC constitutes the other extensive, open-source dataset of scientific documents \cite{lo2019s2orc}. Many other collections that incorporate scientific papers rely on S2ORC for their scientific collections, including the Dolma \cite{soldaini_dolma_2024_correct_caps} and the RedPajama family of datasets \cite{elazar2023s}.

Since data curation is crucial for training ever-larger language models, it is likely that leading companies such as OpenAI, Meta, Google, and Mistral have developed proprietary parsing tools to handle this task. Microsoft's Donut and Meta's Nougat demonstrate their capability to do so. Nevertheless, specific tools, document collections, and computational scales remain undisclosed.

\section{Problem Statement}\label{sec:problem}
An ideal parsing strategy will maximize the accuracy of text output while minimizing the computational cost of obtaining it. We formalize this notion to design AdaParse that optimally balances accuracy and runtime considerations.

\subsection{Accuracy}
Consider $d_{i}$ to be a PDF document that is identified by an index $i \in [n]$ and spans pages of text, tables, and figures. Denote the associated (groundtruth) text by $\psi_{i} = \psi(d_{i}) \in \Sigma^{\star}$, a sequence of characters over an alphabet $\Sigma$ (e.g., Unicode). While the alphabet is usually known, the document's groundtruth text is not directly observable and must be approximated by a parser.

There is a set of parsers $\left\{ \phi_{1}, \ldots , \phi_{m} \right\}$ available to retrieve text from a document collection $\{d_{1},...,d_{n}\}$. Invoking a parser $\phi_{j}$ on a PDF document $d_{i}$ provides an approximation of its groundtruth text $\phi_{j}(d_{i}) \approx \psi_{i}$. The quality of the approximation can be assessed by an accuracy metric $\mathcal{A}$ that may be defined by
$$\mathcal{A} \left( \phi_{j}, d_{i} \right) = a \left( \lVert \phi_{j}(d_{i}) - \psi_{i} \rVert_{\Sigma^{\star}} \right)$$
through some norm over the alphabet $\lVert \cdot \rVert_{\Sigma^{\star}}$ and a monotonically decreasing function $a$ mapping the distance of the strings to a quality score. The accuracy measure is abstract not because of mathematical convenience but due to the nature of document parsing: It is unclear what type of dissimilarity best captures scientifically sound and faithful parser text output. 

The computational resources (e.g., runtime) required to parse a document are given by $\mathcal{T}_{k} \left( \phi, d \right)$. The resource usage for $k \in \{ \emph{CPU}_{\textrm{mem}}, \emph{CPU}_{\textrm{time}}, \emph{GPU}_{\textrm{mem}}, \emph{GPU}_{\textrm{time}} \}$ depends on the document, parser, and system used.

We formalize the trade-off between accuracy and efficiency by assigning any of the $m$ parsers to each of the $n$ documents individually, i.e., $j_i \in [m]$, and optimize the following conflicting objectives. For any assignment of parsers to the dataset of documents $\mathbf{j} = (j_1, \dots, j_n) \in [m]^n$ we want to maximize overall accuracy
$$\max_{\mathbf{j}} \left\{ \sum_{i=1}^{n} \mathcal{A} \left( \phi_{j_i}, d_i \right) \right\}$$
while simultaneously minimizing total computational cost
$$\min_{\mathbf{j}} \left\{ \sum_{i=1}^{n} \mathcal{T}_{k} \left( \phi_{j_i}, d_i \right) \right\} \mbox{.}$$

Imposing a constraint on one objective while optimizing the other strikes a balance. Since accuracy is not observable, we aim to maximize its conditional expectation by selecting the appropriate parser $\phi_{j_{i}}$. The expectation is conditioned on the document $d_{i}$'s first page's text $\phi^{1}_{1}(d_{i})$ parsed by the default parser $\phi_{1}$. The resulting constrained optimization problem 
\begin{align*}
\max_{\mathbf{j}} & \left\{ \sum_{i=1}^{n} \mathbb{E} \left[ \mathcal{A} \left( \phi_{j_i}, \psi_i \right) \, | \, \phi_{1}^{1}(d_{i}) \right] \right\} \\
\qquad {s.t.} & \quad \sum_{i=1}^{n} \mathcal{T} \left( \phi_{j_i}, d_i \right) \leq \overline{\mathcal{T}}
\end{align*}
conveys a crucial property. We can partition the dataset into subsets of $\{n_{1},...,n_{L}\}$ documents such that $\sum_{i=1}^{L} n_{i}=n$ and process them across $L$ nodes. If each node $l \in [L]$ adheres to a computational budget of $\frac{n_{l}}{n}\overline{\mathcal{T}}$, the overall required resources will not exceed $\overline{\mathcal{T}}$. Therefore, adaptive document parsing with heterogeneous parsing algorithms can be realized through embarrassingly parallel workloads. A tuning parameter $\alpha \in [0,1]$ controls the trade-off between (expected) accuracy and runtime in AdaParse.

\subsection{Direct Preference Optimization}
While parsing accuracy and its trade-off with efficiency appear vague, scientists usually have a strong preference when they are faced with different parser outputs of the same document $\phi_{1}(d_{i})$ and $\phi_{2}(d_{i})$, irrespective of whether groundtruth text $\psi_{i}$ is available. Therefore, instead of fixing an accuracy measure to $\mathcal{A}=\mbox{ROUGE}$, for example, we attempt to (implicitly) learn one through user preferences. This is not far-fetched, as accuracy measures like BLEU or ROUGE are designed to be strongly correlated with human preferences \cite{reiter2018structured}. Since a (predicted) accuracy measure only serves as a means to assign a parser to a document, we allow the predictive model to learn this assignment directly from user input through direct preference optimization (DPO).

Connecting user preference with predicted accuracy has a practical rationale. Malformed text in the parser output $\phi_{j}(d_{i})$ is indicative of overall parser quality. Moreover, there are specific patterns that strongly inform accuracy estimates and human perception of quality alike; see \autoref{fig:failuremodes}. Training a model to infer accuracy from the presence of such malformed text patterns offers a foothold to learn to select a parser adaptively.

In principle, a model capable of assigning a scalar to text, i.e., $\pi_{\theta}: \Sigma^{\star} \rightarrow [0,1]$, is a potential candidate for inferring (normalized) accuracy. Rule-based approaches or classical ML models offer interpretable and tractable solutions. Expressive models such as LLMs, on the other hand, that were pre-trained on broad textual data, can be fine-tuned in text sequence regression to make a prediction on text accuracy based on subtle features.

Given a sufficient dataset, a model $\pi_{\theta}$ can be fine-tuned to predict the BLEU accuracy. This is the crucial ingredient to allow a parsing strategy to predict a good parser-document matching. Moreover, recent strategies for aligning LLMs with human preferences can allow such a model to infer an accuracy measure (implicit in the model). 

\section{Design}\label{sec:design}

\subsection{Overview: Why Adaptive Parsing}
The empirical results indicate that the versatility of the layout and textual content of PDF documents prevents assigning a likely parser through deterministic rules alone. For example, the scientific category to which a document belongs (as indicated by associated keywords) is only a weak indicator of its actual content and the difficulty of parsing it. For instance, a research paper on machine learning may boast hundreds of LaTeX expressions, more akin to a mathematics paper. Similarly, document metadata such as the publication year can fail to represent the quality of the embedded text, as that text may have been attached with state-of-the-art OCR software long after the document's publication. Regardless, obtaining any of these features requires parsing the document. In turn, choosing the optimal parser for a document appears to require parsing it beforehand. 

We cut this Gordian knot by leveraging text extraction to inform if and what subsequent text recognition algorithm (OCR or ViT) should be run. In particular, PyMuPDF offers exceedingly fast text extraction, with a throughput 135$\times$ higher than Nougat and 13$\times$ greater than that of pypdf. Thus prefacing parser selection with it is computationally cheap. Furthermore, the lower accuracy of PyMuPDF works partially in our favor: Malformed substrings (e.g., of LaTeX equations, whitespace, or scrambled characters) that are typical of text extraction output are informative for the predictive parser selection algorithm. 

\begin{figure}[h]
    \centering
    \includegraphics[width=0.75\columnwidth]{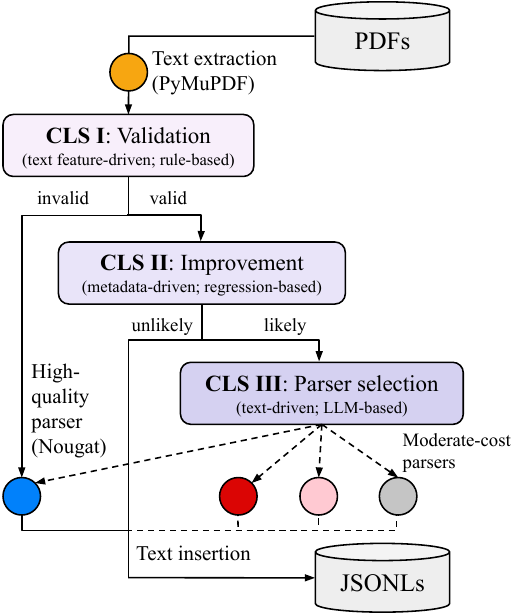}
    \caption{System architecture diagram for a range of predictive models: After an initial text extraction step (PyMuPDF), PDFs are routed through a hierarchical classification pipeline. \textbf{CLS I} predicts the binary quality attribute of the extracted text through coarse but fast-to-compute features (e.g., text length). For valid texts, \textbf{CLS II} assesses if an improvement is likely for any other parser. If affirmative, \textbf{CLS III} selects the parser most likely to improve output text quality.}
    \label{fig:architecture}
\end{figure}

Parser selection can be performed as a hierarchical classification scheme based on the PyMuPDF-extracted text. The first classification stage CSL~I employs aggregate statistics computed from the extracted text (e.g., number of characters) to infer validity. While simplistic, the features are highly interpretable and permit rapid inference. If the PyMuPDF text is deemed invalid, the PDF is sent to the high-quality Nougat parser. If, however, the text is deemed valid, a second classification stage CSL~II is applied to determine if parsing with another parser (including Nougat) may nevertheless bring a significant improvement in parse quality. This binary label is inferred from metadata (e.g., authoring tool, year of publication, number of pages). If a significant improvement is deemed unlikely, the PyMuPDF-extracted text is accepted as the document parse and is subsequently written to storage. On the other hand, if improvement is predicted as likely, the third classification stage CSL~III is applied to select the parser. Since this decision is based on subtle patterns in the extracted text, a fine-tuned LLM is invoked for this multi-class downstream task.

We introduce two implementations. The first variant, \textbf{AdaParse (FT)}, implements the classification stages CLS~I and CLS~II within a single routine. If an improvement appears likely, it directly triggers Nougat rather than weighing its options with other moderate-cost parsers. Therefore, it does not invoke an LLM and skips stage CLS~III. It employs predefined fastText (FT) word embeddings \cite{xu2019deep}.

\begin{figure}
    \centering
    \includegraphics[width=\columnwidth,trim=13mm 4mm 15mm 12mm,clip]{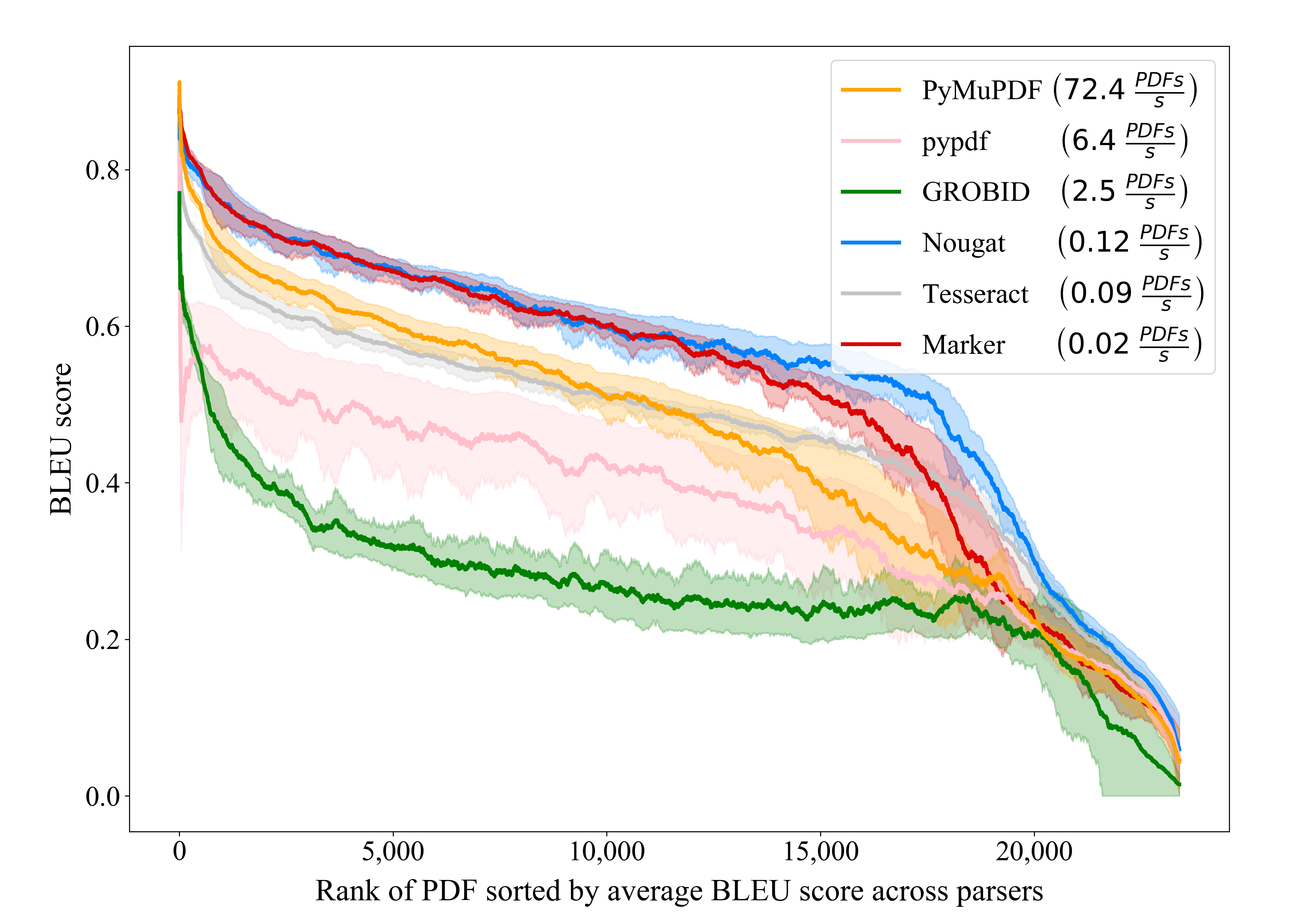}
    \caption{Parser performance (BLEU) for $n$ = 23,398 PDFs. They are sorted by parsing difficulty which is estimated for each document by the average BLEU score across parsers. The higher the rank, the greater the estimated parsing difficulty. Throughputs for a single node using each parser are presented in the legend.}
    \label{fig:motivateadaptive}
\end{figure}

The second variant, \textbf{AdaParse (LLM)}, implements the first classification stage, CLS~I, to determine if the extracted text is worthy of being included in a batch and run through LLM inference. Once a batch of text items is assembled, this variant proceeds directly to the stage CLS~III, which performs an LLM inference call to predict the most suitable parser for each text item. We employ SciBERT \cite{beltagy2019scibert} for this task due to its high inference speed. Consequently, the single‑node throughput is still 17$\times$ higher than that achieved by solely relying on a state-of-the-art ViT-based parser (Nougat). We find that this use of LLM inference results in slightly lower throughput than the first variant---although still matching the performance of a text extraction tool such as pypdf---but offers higher accuracy and allows for better alignment with human preferences through DPO. The LLM-inferred labels determine if the extracted text is accepted as is or if the document (still in memory) is routed to a high-quality parser such as Nougat. 

In essence, AdaParse is a meta‑strategy that adaptively ensembles multiple parsers into a single, higher‑accuracy system—loosely inspired by AdaBoost \cite{freund1996experiments}. We employ Parsl \cite{babuji19parsl}, a pure-Python parallel scripting library, to orchestrate AdaParse's data processing and model inference on the Polaris supercomputer.

\subsection{Optimizing Parallel Execution on HPC Systems}

Nougat serves as the high-quality parser in AdaParse. In the following, we restrict its usage to (at most) $\alpha=5\%$ of the documents per node. Regardless, invoking Nougat requires loading a Swin-architecture-based Vision Transformer—which can take up to 15 seconds on an A100.
Thus, we modify Parsl to allow Nougat to persist on each GPU beyond the task boundary. Since PyMuPDF, the best lightweight parser, runs exclusively on CPUs, there is effectively no competition with Nougat for GPUs, allowing for efficient resource sharing.

\begin{figure}
    \centering
    \includegraphics[width=\columnwidth]{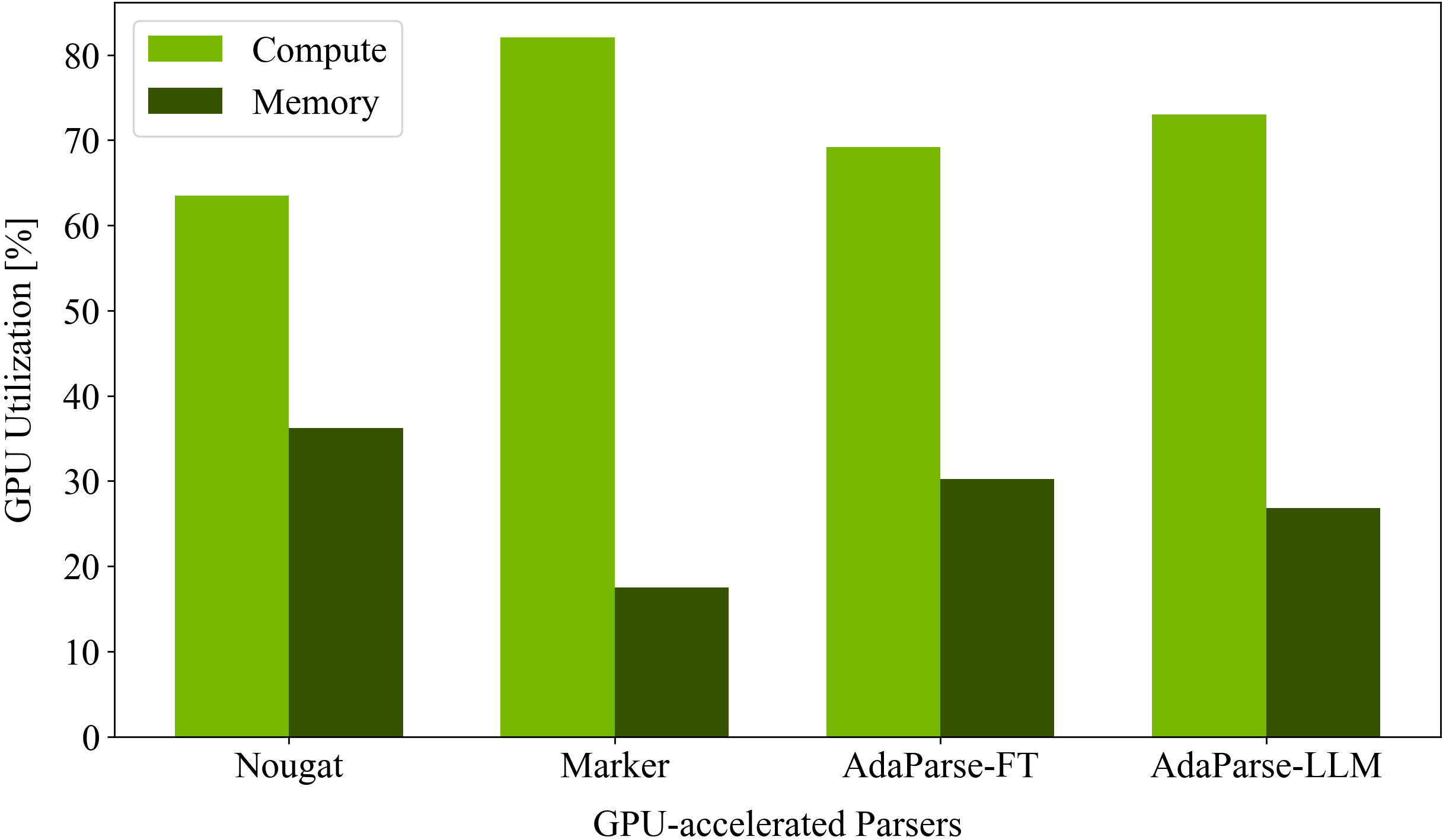}
    \caption{Utilization of the workload per GPU, as measured with the NVIDIA Nsight Systems profiler (Nsys).}
    \label{fig:gpuusage}
\end{figure}

Nougat operates on a fixed image input size of (\textit{H},\textit{W}) = (896, 672), but allows control over how many pages are processed simultaneously. We find that a batch size of $B_{p}$=10 pages maximizes throughput without exceeding GPU memory capacity. Although Nougat’s Base model is relatively small (350M parameters), its memory footprint grows substantially when document pages are converted into image patches for the self‑attention mechanism. By parsing pages individually at a fixed resolution—rather than entire documents—Nougat normalizes task size, resulting in more consistent execution times.

The performance analysis of the GPU-accelerated parsing methods was conducted using the NVIDIA Nsight Systems profiler (Nsys), and the results are presented in \autoref{fig:gpuusage}.

%

\section{Experimental Methodology}\label{sec:methods}
\subsection{Hardware and Software Environment}

All experiments were conducted on the Polaris system at the Argonne Leadership Computing Facility (ALCF). Polaris is an HPE Apollo Gen10+ system with 560 nodes interconnected by an HPE Slingshot-11 network with a Dragonfly
topology. Each node consists of an AMD “Milan” processor with 32~cores and 512~GB of system memory, four 40~GB NVIDIA A100 GPUs,
and two Slingshot-11 25~GB/s network adapters. Each NVIDIA A100
GPU can achieve a peak of 19.5~TFLOPS in FP32 and 312 TFLOPS in FP16 and BF16. Polaris is supported by a Lustre file system, Eagle, residing on an HPE ClusterStor E1000 platform equipped with 100~PB of usable capacity across 8480 disk drives. This ClusterStor platform also provides 160 Object Storage Targets (OST) and 40 Metadata Targets (MT) with an aggregate data transfer rate of 650 GB/s. All experiment data is striped across 48 OSTs for optimal read and write bandwidth.
As a current top 30 supercomputer, Polaris is representative of leadership class HPC systems.

We employ the Parsl workflow engine to orchestrate our PDF parsing effort efficiently. Parsl distributes tasks as pure functions---deterministic operations that do not modify shared program state---which poses challenges when the same ML model weights are needed across hundreds of workers pinned to GPUs. To mitigate this problem, we implement a warm-start mechanism for parsers requiring machine learning models. By loading the model weights once and persisting them across worker processes, we significantly reduce I/O overhead and initialization time for subsequent tasks. Furthermore, to decrease global I/O usage, we aggregate and chunk input files into a set of compressed ZIP archives and transfer them to node-local RAM storage. This strategy minimizes the frequent reading and writing of numerous small files to networked file systems, instead favoring larger, more efficient I/O operations suited to Lustre file systems. By processing data locally on each node, we enhance throughput and reduce the load on shared storage resources. Parsl dispatches tasks adaptively based on worker availability, ensuring efficient use of compute resources by dynamically balancing task distribution across nodes.

\subsection{Document Selection and Preparation}

We employ a diverse set of documents and formats, from both preprint servers and peer‑reviewed publishers, for evaluating the parsers to ensure they can capture the diversity of scientific text. Diverse sources are important because different venues use different templates and represent different levels of polish in scientific works.
The dataset includes documents sourced from ArXiv, BioRxiv, BMC, MDPI, MedRxiv, and Nature.
The resulting collection spans eight domains (mathematics, biology, chemistry, physics, engineering, medicine, economics, and computer science) with 67 sub-categories ranging from acoustics to zoology.
Including such a wide range of topics is critical to obtaining a comprehensive representation of different domain-specific features, such as extensive use of equations in mathematics and differing notations, conventions, citation schemes, and formatting used in various fields \cite{shah2021math}. 

We focus on recent data that would not have been available for ViT/OCR models to train on, in order to prevent data leakage from their training set into our test set.
This choice presents a trade-off: it excludes older documents, which may contain metadata of varying quality for extraction tools like PyMuPDF.

To obtain groundtruth text for the benchmark, we parsed the HTML representation of a paper's full text allowing us to obtain a sufficiently large number of documents for evaluation. Since HTML is straightforward to parse, it provides highly accurate groundtruth text.

Lastly, we perform page image and text layer manipulations (e.g., random scaling, artificial image imperfections, or modified metadata), as done in prior works \cite{zi2005groundtruth,groleau2023augraphy}.
This approach ensures that we evaluate the parsers as they would be applied ``in the wild,'' to obtain results that are representative of real-world performance.

\subsection{User Preferences}
Aligning accuracy with human preferences requires sampling those preferences. For this purpose, we launched a platform that allows domain experts to share their preferences on texts sourced from seven different parsers. 
The expert is presented with an image of a document page along with two parsed text outputs, and is prompted to either choose a preferred parse or indicate indifference if neither is preferred.
The text formatting of the parser output was slightly modified to prevent bias (e.g., by including or removing hashtags that indicate markdown output from Nougat or Marker). Moreover, the selection of page and text pairs was non-adaptive to prevent user feedback bias \cite{mansoury2020feedback}. To ease users into this task, the website's design emulates that of an OpenAI chatbot \cite{chiang2024chatbot}. Moreover, users began annotating single paragraphs before moving on to entire document pages. 

We engaged 23 scientists with expertise spanning mathematics, biology, physics, chemistry, medicine, engineering, and economics. We obtained 2794 preferences for 642 different document pages, which we partitioned into training, validation, and test subsets with sizes 712, 234, and 1848, respectively. The majority of the preferences were collected for the test subset to a) ensure a sufficient sample size to uphold the validity of the empirical results and b) present identical options to different users to assess consensus.

\section{Evaluation}\label{sec:evaluation}
After outlining how data was sourced and models configured, we present our results.

\subsection{Alignment of Accuracy with User Preferences}
We first need to assess if BLEU scores and similar metrics are good proxies for human preferences.
To do this we study the outcomes of our user preference survey.

When aggregating over the entire dataset, users prefer Nougat the most (with a frequency of 57.1\%) followed by Marker (49.1\%) and PyMuPDF (48.6\%). Throughput does not necessarily translate to user preferences. PyMuPDF, for example, offered a 2133$\times$ higher throughput while experiencing a BLEU score difference of 0.5\%. However, users are not indifferent to parsers, as indicated by the low win frequency of 2.1\% for pypdf. As these frequencies are determined by a binary tournament of different pairings of the seven parsers present in the study, percentages do not sum to 100\%. Therefore, we report normalized win rates instead. 

Users are highly willing to make their preferences known, doing so 91.3\% of the time and while picking "neither" only in the remaining 8.7\%. Moreover, participants have a high agreement in the choices they make. Among the 405 triplets of page document and two parser output texts shown to multiple users, participants made the same choice 82.2\% of the time. This high consensus rate—achieved despite scientists’ diverse disciplinary backgrounds—suggests a degree of objectivity in participants’ preferences, underscoring their usefulness for model refinement. Importantly, these preferences are collected only once and used offline to adapt the model’s weights via DPO during post‑training, so that no further human input is required when the model is deployed for parsing.

A key result of this study is that the BLEU score, while indicative of user preference, is hardly predictive. The BLEU is highly correlated with the win rate (correlation $\hat{\rho}$ = 0.47), which is statistically significant as $H_{0}:\rho=0$ is rejected with $p$ = 8.4\textsuperscript{--49}. Yet the correlation is also far from 1, explaining only 47\% of the variation in user choices. We view this result as justification that the BLEU score is a robust quality indicator of parser text output and a suitable target for LLM-finetuning, but also not completely predictive requiring the consideration of other measures of quality.

\subsection{AdaParse Quality Assessment}
Since AdaParse manages diverse parsers during its execution, it is important to probe the mechanism for parser selection and to gauge the improvement over using individual parsers.
Because no single quality measure is completely predictive of user preferences, we consider a set of quality measures.
The empirical investigation includes document coverage (as measured by the number of retrieved document pages), BLEU, ROUGE, and character-accuracy rate (CAR). It also includes two metrics we devised from the user preference study: \textit{win rate} (WR) which measures how often a parser was selected over the others for a given document and \textit{accepted tokens} (AT) that tracks the relative frequency of tokens that exceed a critical BLEU threshold.

To evaluate AdaParse, we run it on a held-out test set of 1000 digitally born PDFs that were not used during the training. We report three rounds of metrics: the first with no changes to any layer of the PDFs, the second with augmentations applied only to the image layer, and the third with alterations applied only to the text layer.

We show in Table~\ref{tab:unmodified} the default quality on the test set. Marker has the highest coverage rate, but does not have the highest quality according to any other metric. Nougat has the highest win rate between parsers by a slim margin. AdaParse even with the requirement to allocate no more than 5\% of the documents to its high-quality parser (Nougat), produces the best BLEU and ROUGE scores, and the second-best CAR. Additionally, AdaParse has the highest percentage of accepted tokens based on the user preference data at 76.9\%.
AdaParse can achieve better performance than any of its constituent parsers by delegating to the method that is most suitable for an individual document. While a parser like Nougat may perform best on average, it is not the best parser for each document allowing AdaParse to exceed it if it accurately infers a better parser-document matching.

The performance of AdaParse is based on the capability in predicting the BLEU score of PyMuPDF and Nougat-parsed text, with an $R^{2}=40.0$\% and $R^{2}=46.5$\%, respectively. This is largely based on parameter-efficient finetuning through low-rank adaptation (LoRA) \cite{hu2021lora} and DPO on its weights in decoder mode. DPO post‑training has been shown to improve performance in related downstream prediction tasks using relatively little preference data, even in high‑performance computing (HPC) applications \cite{dharuman2024mprot}.

\begin{table}[htbp]
    \caption{Accuracy on born‑digital PDFs: Document- (coverage rate), word- (BLEU, ROUGE), and character-level (CAR) accuracies. CAR = Character accuracy rate. WR = Win rate. AT = Accepted tokens. All \%.}
    \begin{center}
    \resizebox{\columnwidth}{!}{%
    \begin{tabular}{|c|c|c|c|c|c|c|}
    \hline
    \textbf{Parser} & \textbf{Coverage} & \textbf{BLEU} & \textbf{ROUGE} & \textbf{CAR} & \textbf{WR} & \textbf{AT} \\
    \hline
    Marker & \textbf{96.7} & 47.5 & 64.2 & 59.6 & 26.6 & 73.3 \\
    \hline
    Nougat & 93.0 & 48.1 & 66.5 & 65.8 & \textbf{27.9} & 69.8 \\
    \hline
    PyMuPDF & 91.3 & 51.9 & 67.3 & 67.0 & 24.4 & 76.7 \\
    \hline
    pypdf & 92.0 & 43.6 & 58.7 & 32.3 & 2.4 & 72.4 \\
    \hline
    GROBID & 81.0 & 26.5 & 52.4 & 54.8 & -- & 20.6 \\
    \hline
    Tesseract & 91.3 & 48.8 & 64.2 & \textbf{67.8} & 18.7 & 72.5 \\
    \hline
    AdaParse & 91.5 & \textbf{52.1} & \textbf{67.6} & 67.1 & 25.5 & \textbf{76.9} \\
    \hline
    \end{tabular}%
    }
    \label{tab:unmodified}
    \end{center}
\end{table}

Additionally, we test parsing performance under simulated image degradation to mimic low-quality scans.
Low-quality scans are common in older academic and book datasets.
We emulate this quality degradation with random rotations, contrast adjustments, Gaussian blurring, and compression that are applied to a subset of 15\% of documents, similar to the data augmentations used to train Nougat \cite{blecher2023nougat_new}. Note that these changes will not affect text extraction methods which is why we exclude them here. AdaParse shows favorable performance as it relies mostly on text extraction that is unaffected by these changes and Nougat that was trained to handle similar image augmentations. The only statistically meaningfully affected metric is the win rate---in part because AdaParse is artificially limited to selecting an image parser when its quality is higher. However, it is important to note that this does not translate to a lower token acceptance rate, as many documents are still parsed above the acceptance threshold.

\begin{table}[htbp]
    \caption{Accuracy on simulated scanned PDFs: Document- (coverage rate), word- (BLEU, ROUGE), and character-level (CAR) accuracies. All \%.}
    \begin{center}
    \resizebox{\columnwidth}{!}{%
    \begin{tabular}{|c|c|c|c|c|c|c|}
    \hline
    \textbf{Parser} & \textbf{Coverage} & \textbf{BLEU} & \textbf{ROUGE} & \textbf{CAR} & \textbf{WR} & \textbf{AT} \\
    \hline
    Marker & \textbf{96.5} & 46.6 & 62.9 & 60.5 & \textbf{28.0} & 70.1 \\
    \hline
    Nougat & 91.9 & 45.1 & 63.1 & 63.4 & 27.2 & 63.5 \\
    \hline
    Tesseract & 90.0 & 44.0 & 58.2 & 65.2 & 12.8 & 59.0 \\
    \hline
    AdaParse & 92.8 & \textbf{52.0} & \textbf{67.5} & \textbf{67.0} & 18.4 & \textbf{77.0} \\
    \hline
    \end{tabular}%
    }
    \label{tab:impact_image_manipulation}
    \end{center}
\end{table}

Finally, we investigate the perturbation of the text layer. 15\% of the embedded text layers are replaced with the output of common tools (Tesseract or GROBID), explaining their exclusion in the table. This configuration tests the ability to determine when a higher-quality parse is needed because of degraded text.
Few documents parsed by Nougat are sufficient to give AdaParse an edge in this setting. 
Given that we still limit AdaParse's choice of image parsers to at most 5\% of the dataset compared to the 15\% where the text layer is removed, it is unsurprising that quality degrades commensurate with the text parsers that are being used for the bulk of the parsing efforts. Regardless, AdaParse correctly delegates sufficiently many of those documents to other parsers which is why quality remains higher than using text extraction-based parsers alone.

Overall, AdaParse offers robust performance across data regimes. 

\begin{table}[htbp]
    \caption{Accuracy on PDFs with simulated OCR‑degraded text layers: Document- (coverage rate), word- (BLEU, ROUGE), and character-level (CAR) accuracies. All \%.}
    \begin{center}
    \resizebox{\columnwidth}{!}{%
    \begin{tabular}{|c|c|c|c|c|c|c|}
    \hline
    \textbf{Parser} & \textbf{Coverage} & \textbf{BLEU} & \textbf{ROUGE} & \textbf{CAR} & \textbf{WR} & \textbf{AT} \\
    \hline
    PyMuPDF & 90.8 & 42.0 & 55.6 & 56.5 & \textbf{13.1} & 58.8 \\
    \hline
    pypdf & \textbf{91.2} & 35.6 & 48.9 & 29.8 & 1.2 & 56.9 \\
    \hline
    AdaParse & \textbf{91.2} & \textbf{42.4} & \textbf{55.9} & \textbf{56.7} & 12.0 & \textbf{59.5} \\
    \hline
    \end{tabular}%
    }
    \label{tab:textmanip}
    \end{center}
\end{table}

\subsection{Throughput Scalability}
In addition to quality, throughput is the key evaluation criterion for large-scale parsing techniques. If you cannot perform a parse with a given parser due to insufficient resources, the quality of that parser becomes moot.

\begin{figure}
    \centering
    \includegraphics[width=\columnwidth]{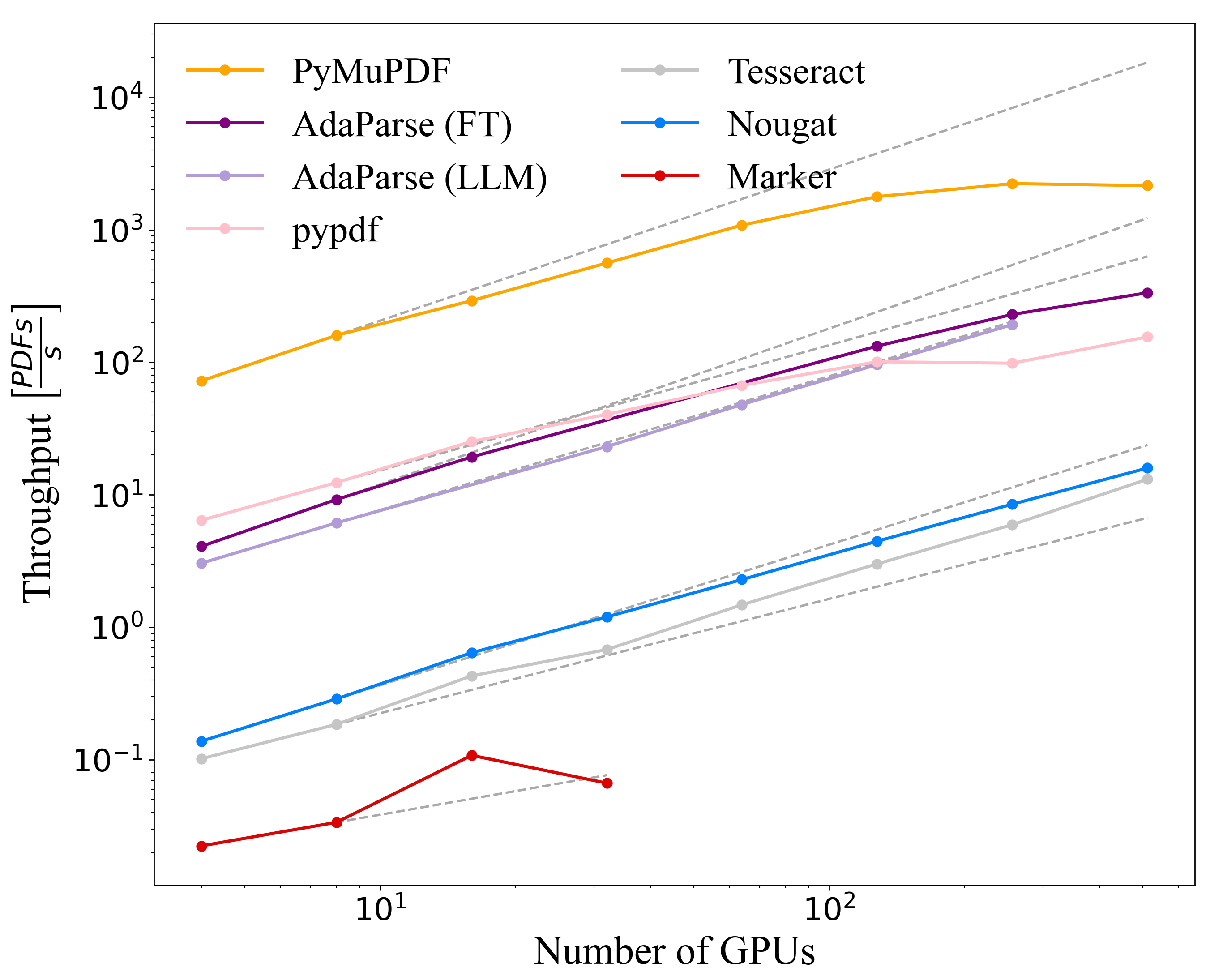}
    \caption{Scalability of the seven parsers.}
    \label{fig:scalability}
\end{figure}

We evaluate the performance of each method on Polaris using between 1 and 128 nodes and present the results in \autoref{fig:scalability}.
We see that text extraction-based methods such as PyMuPDF are the fastest, processing up to $\approx$315 PDF/second at scale, while methods like Marker fail to scale beyond 10 nodes, producing on average only 0.1 PDF/second.  Nougat offers slightly better throughput with $\approx$8 PDF/second on 128 nodes.
AdaParse (FT) exhibits intermediate performance with $\approx$78 PDF/second.
Most methods scale roughly linearly in the number of nodes. Notable exceptions include PyMuPDF and pypdf, which initially scale linearly but plateau at around 128 and 100 nodes, respectively.  In the case of PyMuPDF, plateauing occurs because extraction is sufficiently fast that contention for file system resources begins to be the bottleneck, limiting scalability.

\section{Conclusion}\label{sec:conclusions}
A key step in extracting knowledge from scientific documents is to improve the quality of PDF document parsing. This step has long been a bottleneck for building AI foundation models for science. While tools developed for Internet-scale data are widely adopted for training advanced AI models, training on text encoded within scientific literature data remains an open challenge. We have presented AdaParse as a practical approach to address this challenge---mainly by incorporating a data-driven strategy to distribute each parsing task to the most appropriate PDF parser tool, in a portable, yet reusable manner. We also developed direct preference optimization for rating PDF parser quality, allowing selections to be aligned with human judgment. We show that the resulting solution is able to leverage existing PDF parsing tools for large-scale campaigns that make effective use of high-performance computing infrastructure.


\section*{Acknowledgments}
This research used resources of the Argonne Leadership Computing Facility, a U.S.\ Department of Energy (DOE) Office of Science user facility at Argonne National Laboratory (ANL) and is based on research supported by the DOE Office of Science--Advanced Scientific Computing Research Program and by Laboratory Directed Research and Development (LDRD) funding from ANL, provided by the Director, DOE Office of Science, both under Contract No.\ DE-AC02-06CH11357.


\newpage
\bibliographystyle{mlsys2025}
\bibliography{bibtex/bib/bare_jrnl}

\onecolumn
\newpage
\twocolumn

\appendix

\section{The DPO Formalism}
We denote the parsed text of document $i$ by parser $j$ as $x_{i}^{j} = \phi_{j}(d_{i})$ with accuracy (e.g., BLEU score) $y_{i}^{j}$. Hence, the dataset $\mathcal{D} = \{(x_i, y_i)\}_{i=1}^N$ represents the (parsed) text inputs with $\mathbb{R}^{m}$-valued responses (i.e., a document-wise accuracy vector). We post-train a model to predict the accuracies of all parsers given the default parser's text $\phi^{1}_{1}$ in three steps. First, supervised fine-tuning yields the estimate $\hat{\theta}_{1}$ through minimization of the $\ell_{2}$ loss
$$\mathcal{L}_{\text{REG}}(\theta) 
= \mathbb{E}_{\mathcal{D}} \left[ \lVert \pi_{\theta}(x^{1}) - y \rVert_{2}^{2} \right] \mbox{.}$$
Second, $\pi_{\hat{\theta}_{1}}$ is augmented into an encoder-decoder model $g_{\varphi}$, with $\text{Enc}_{\varphi_e}(x)=h$ and $\text{Dec}_{\varphi_d}(h) = z$, where $\varphi = (\varphi_e, \varphi_d)$ and $\varphi_e := \hat{\theta}_{1}$ initially. We utilize a preference dataset $\mathcal{D}_{\text{pref}} = \{( x_{j}^{k_{1},+}, x_{j}^{k_{2},-} ) \}_{j=1}^M$ of text pairs obtained through different parsers $\phi_{k_{1}}$ and $\phi_{k_{2}}$ where the former is preferred by the user. Minimizing
$$\mathcal{L}_{\text{DPO}} = - \mathbb{E}_{\mathcal{D}_{\text{pref}}} \left[ \log \sigma \left( \theta \log  
\frac{g_\varphi(x^{+})}{g_{\varphi}^{\text{ref}}(x^{+})} -
\log \frac{g_\varphi(x^{-})}{g_{\varphi}^{\text{ref}}(x^{-})}
\right) \right]$$
upon convergence yields the estimate $\hat{\theta}_{2} = \hat{\varphi}_{e}$. Finally, the updated encoder is fine-tuned on $\mathcal{D}$ with a lowered learning rate to obtain $\hat{\theta}_{3}$ which produces the final model.

In our setting, the regression dataset contains N=29,200 pairs, each consisting of a single document text and its associated BLEU score. The output dimension is m=6, since we predict the accuracy for each parser. The preference dataset contains M=712 pairs. We found it advantageous in step 1 to predict pagewise accuracy (i.e., predict the accuracy of the given page's parsed text), while the regression data in the third step are used to infer document-level accuracy based on the first page's text, as processed by AdaParse.

\section{Quantification of the DPO Impact}
We quantify the benefit of direct preference optimization (DPO) by evaluating a range of prediction models. As a baseline, we apply support vector classification (SVC) to metadata features (e.g., publisher, year of publication, PDF format, and producer). LLM-based prediction of the document text is performed with SciBERT, BERT, MiniLM, and SPECTER \cite{cohan2020specter, wang2020minilm}. The metrics of the reference models (BLEU-maximal/minimal and random selection) are provided for context.
\begin{table}[htbp]
    \caption{Evaluation of various prediction models across different features. Word-level (BLEU, ROUGE) and character level accuracy (CAR) accuracies. WR=Win rate. All \%.}
    \begin{center}
    \resizebox{\columnwidth}{!}{%
    \begin{tabular}{|l|c|c|c|c|c|}
    \hline
    Features (Model) & \textbf{BLEU} & \textbf{ROUGE} & \textbf{CAR} & \textbf{WR} & \textbf{ACC} \\
    \hline
\multicolumn{6}{|l|}{\textbf{CLS III: Document Text}} \\ \hline
Text (SciBERT + DPO) & \textbf{52.7} & 69.4 & \textbf{68.0} & \textbf{31.4} & 36.7 \\ \hline
Text (SciBERT) & 51.6 & \textbf{69.5} & 66.9 & 25.0	 & \textbf{48.3} \\ \hline
Text (BERT) & 49.7 & 66.0 & 63.4 & 24.8 & 40.0 \\ \hline
\multicolumn{6}{|l|}{\textbf{CLS II: Metadata and Title Text}} \\ \hline
Title + Metadata (SPECTER) & 47.9 & 64.5 & 62.9 & 25.2 & 18.1 \\ \hline
Title (SPECTER) & 46.4 & 63.3 & 61.8 & 26.2 & 15.2 \\ \hline
Title + Metadata (MiniLM-L6) & 44.7 & 62.2 & 60.4 & 28.4 & 10.1 \\ \hline
\multicolumn{6}{|l|}{\textbf{CLS I: Metadata}} \\ \hline
Format + Producer (SVC) & 47.7 & 64.0 & 60.2 & 28.5 & 14.6 \\ \hline
Format (SVC) & 47.5 & 64.1 & 60.7 & 29.5 & 16.6 \\ \hline
Year + Producer (SVC) & 47.3 & 63.7 & 60.1 & 28.8 & 14.8 \\ \hline
Publisher + (Sub-)category (SVC) & 46.4 & 63.7 & 60.9 & 21.7 & 14.8 \\ \hline
(Sub-)category (SVC) & 43.6 & 63.5 & 62.5 & 24.9 & 12.9 \\ 
\hline
\multicolumn{6}{|l|}{\textbf{\textcolor{gray!90}{Reference}}} \\ \hline
\textcolor{gray!90}{BLEU-maximal selection} & \textcolor{gray!90}{56.8} & \textcolor{gray!90}{72.3} & \textcolor{gray!90}{70.4} & \textcolor{gray!90}{26.5} & \textcolor{gray!90}{100.0} \\ \hline
\textcolor{gray!90}{Random selection} & \textcolor{gray!90}{44.0} & \textcolor{gray!90}{61.7} & \textcolor{gray!90}{57.4} & \textcolor{gray!90}{20.5} & \textcolor{gray!90}{16.7} \\ \hline
\textcolor{gray!90}{BLEU-minimal selection} & \textcolor{gray!90}{21.5} & \textcolor{gray!90}{44.2} & \textcolor{gray!90}{44.6} & \textcolor{gray!90}{18.1} & \textcolor{gray!90}{0.0} \\ \hline
    \end{tabular}%
    }
    \label{tab:results}
    \end{center}
\end{table}

Given the six parsers, predicting the optimal choice for any PDF is challenging. The assignment of the BLEU-maximal parser to each document yields a BLEU score of 56.8\%. Although metadata-driven classification delivers (mostly) favorable results, text-driven regression with LLMs outperforms them across all metrics. Post-training through DPO further boosts BLEU, CAR, and win rate. Transformer-based models pre-trained on extensive scientific corpora, such as SciBERT and SPECTER, outperform models trained on conventional web-scale data like BERT and MiniLM-v6. AdaParse (LLM) leverages SciBERT with DPO post-training for parser selection.

\section{Solving the optimization problem}

 For scalability reasons, AdaParse limits itself to two parsers: PyMuPDF and Nougat. The problem turns to picking either $\phi_{\text{Nougat}}$ or $\phi_{\text{PyMuPDF}}$ for any document $d_{i}$. The average computational cost of a parser can be determined from our scaling experiments and is documented in the legend of Figure 3. They are denoted by $\mathcal{T}^{\text{avg}}_{\text{Nougat}}$ and $\mathcal{T}^{\text{avg}}_{\text{Nougat}}$. The parameter $\alpha \in [0,1]$ limits the fraction of documents parsed with Nougat. The constraint 
\begin{align*}
\sum_{i=1}^{n} \mathcal{T} \left( \phi_{j_i}, d_i \right) &\approx \alpha n \left( \mathcal{T}^{avg}_{\text{Nougat}} - \mathcal{T}^{avg}_\text{{PyMuPDF}} \right) + n \mathcal{T}^{avg}_{\text{PyMuPDF}} \\
&\leq \overline{\mathcal{T}}  
\end{align*}
is (approximately) satisfied for any
$$\alpha \leq \frac{\overline{\mathcal{T}} - n \mathcal{T}^{\text{avg}}_{\text{PyMuPDF}}}{n \left( \mathcal{T}^{\text{avg}}_{\text{Nougat}} - \mathcal{T}^{\text{avg}}_{\text{PyMuPDF}} \right)} \mbox{.}$$
The objective function is now maximized when sorting the documents (by expected accuracy improvement of Nougat over PyMuPDF) and allowing the first $\lfloor \alpha n \rfloor$ documents to be parsed by Nougat. AdaParse conducts this on a per-batch basis to further increase throughput (i.e. for a batch of size $k$ at most $\lfloor \alpha k \rfloor$ documents will be parsed by Nougat). While this per-batch approach may yield a suboptimal solution, the optimality gap is negligible as the batch size is large (e.g. $k$=256 in our case). 

\end{document}